\documentclass[preprint,12pt,authoryear]{elsarticle}

\usepackage[utf8]{inputenc} 
\usepackage[english]{babel}
\usepackage[T1]{fontenc}
\usepackage{caption}
\usepackage{graphicx}

\usepackage{colortbl}

\usepackage{algorithm}
\usepackage{algorithmic}
\usepackage{dsfont}
\usepackage{amsmath,amsthm,amssymb}
\newcommand*\diff{\mathop{}\!\mathrm{d}}

\usepackage{geometry}

\usepackage{tikz,fullpage}
\usetikzlibrary{arrows.meta}
\usetikzlibrary{matrix}
\usepackage{pgfplots}
\usetikzlibrary{positioning}
\definecolor {processblue}{cmyk}{0.96,0,0,0}

\usepackage{stmaryrd}

\newtheorem{mydef}{Definition}
\newtheorem{problem}{Problem}
\newtheorem{prop}{Proposition}
\newtheorem{rem}{Remark}

\usepackage{amssymb}
\usepackage{amsthm}

\journal{Statistics and Probability Letters}

\begin{document}

\begin{frontmatter}


 \title{A constructive method to minimize couple matchings}
 \author{Pierre Bertrand}
 \ead{pierre.bertrand@ens-cachan.fr}
 \affiliation{organization={Laboratoire de Probabilités, Statistique et Modélisation},
            addressline={Sorbonne Université}, 
            city={Paris},
            country={France}}
 \author{Michel Broniatowski}
 \ead{michel.broniatowski@sorbonne-universite.fr}
 \affiliation{organization={Laboratoire de Probabilités, Statistique et Modélisation \& CNRS UMR 8001},
            addressline={Sorbonne Université}, 
            city={Paris},
            country={France}}
 \author{Jean-François Marcotorchino}
 \ead{jfmmarco3@gmail.com}
 \affiliation{organization={Laboratoire de Probabilités, Statistique et Modélisation},
            addressline={Sorbonne Université}, 
            city={Paris},
            country={France}}

\begin{abstract}
This paper provides constructive procedures for the indeterminacy coupling between two marginal distributions, an alternative to independence coupling. It also introduces a drawing under indeterminacy into a mixture of three independent couplings. Leveraging on this decomposition it states that indeterminacy optimally reduces couple matchings, minimizing the expected number of equal couples drawn in a row. Besides it is seen that the Janson Vegelius coefficient is nothing but a deviation to indeterminacy and it is shown that it tends to $0$ when the number of modalities increases.
\end{abstract}

\begin{highlights}
\item Logical Indeterminacy minimizes couple matchings
\item Logical Indeterminacy sums up as a mixture of three independent couplings
\item Janson Vegelius correlation coefficient computes a deviation to Logical Indeterminacy
\end{highlights}


\end{frontmatter}



\section{Introduction}
A correlation criterion usually computes a deviation with respect to some equilibrium. Theoretical considerations lead to consider independence and logical indeterminacy as the only two possible "natural" equilibria or discrete coupling functions, as can be seen making use of a work of Csizar \cite{csiszar1991least}, a summarized version of which is expressed in \cite{bertrandadac}. 

A discrete coupling is a function $C$ operating on two discrete marginal laws $\mu = \mu_1\ldots\mu_p$ and $\nu=\nu_1\ldots\nu_q$ and which defines a probability law $\pi$ on the product space:

\begin{equation*}
\pi_{u,v}=C(\mu_u,\nu_v), ~\forall~ 1\le u\le p,~ 1\le v \le q
\end{equation*}

We respectively quote both those above mentioned couplings $C^{\times}$ (independence) and $C^{+}$ (indeterminacy); this last notion has been initially introduced by J.-F. Marcotorchino in his seminal paper~\cite{MAR84}); formulas will be reintroduced later on in section~\ref{sec:transport}.

Their usefulness arises in statistical applications: namely, most of our
usual statistical deviation criteria for contingency analysis are expressed in terms of deviations from one of the two couplings (see \cite{PCThese}, which gathers a classification of such criteria); the famous $\chi^2$ index, widely used in practice, computes a deviation to the independence coupling of the empirical margins. Symmetrically the Janson-Vegelius coefficient, initially introduced in \cite{JV77} as a contingency association index, measures a deviation to indeterminacy; we shall detail this point in section~\ref{sec:jv}.

Indeterminacy appears as a poorly known coupling, whose properties have been rarely presented in an explicit way. Its property in relation to minimization of couple matchings (definition~\ref{def:couple_matching}) that it is useful in the Guessing or in the Task Partitioning problem is noticed in \cite{bertrand2021minimization}. The present paper, is precisely dedicated to the properties implied by indeterminacy: \textit{i)} we recall that indeterminacy aims at minimizing couple matching occurrences \textit{ii)} we estimate the probability for a couple of margins uniformly and independently drawn to be eligible for an indeterminacy coupling (property~\ref{prop: validProportion}) \textit{iii)} we decompose an indeterminacy coupling into a mixture of three independent couplings leading to a constructive drawing; it enables to explain the couple matching minimization (property~\ref{prop:drawing+}) \textit{iv)} we analyze the Janson Vegelius correlation coefficient whose expression is nothing but a deviation to indeterminacy.

The paper is structured as follows. Section~\ref{sec:transport} gathers a summarized version of the construction of indeterminacy. The first part of section~\ref{sec:properties} provides the measure of the space of margins eligible for an indeterminacy coupling. A second part is dedicated to the decomposition of indeterminacy. This decomposition is new, to the best of our knowledge, and conveys an interpretation of the initial formula. Section~\ref{sec:jv} gathers an analysis of the Janson Vegelius coefficient.

\section{Construction of indeterminacy \label{sec:transport}}

Although being the most natural, independence is not, by far, the only existing
available coupling method; actually, as introduced by \cite{Sklar73}, any copula function will lead to a coupling function acting on two cumulative distribution functions. 

In the discrete case, two probability measures $\mu =\mu_1\ldots\mu_p$ and $\nu = \nu_1\ldots\nu_q$ represent the initial margins to be coupled. The first one belongs to the simplex $S_p$ of dimension $p$ while the second belongs to $S_q$ of dimension $q$. A coupling $\pi$ of $\mu$ and $\nu$ is as an element of $S_{pq}$ whose margins are $\mu$ and $\nu$, meaning:

\begin{minipage}{0.4\textwidth}
    \begin{equation}\label{eq:marginnu}
        \sum_{u=1}^p\pi_{u,v} = \nu_v,~\forall 1\le v \le q 
    \end{equation}
\end{minipage}
\hspace{4ex} % eventuellement
\begin{minipage}{0.4\textwidth}
    \begin{equation}\label{eq:marginmu}
	    \sum_{v=1}^q\pi_{u,v} = \mu_u,~\forall 1\le u \le p 
    \end{equation}
\end{minipage}

We quote $\mathcal{L}_{\mu,\nu}$ the subset of $S_{pq}$ whose elements obey Equation~(\ref{eq:marginnu}) and Equation~(\ref{eq:marginmu}). It defines the space of couplings of $\mu$ and $\nu$.

\subsection{Reducing the information conveyed by the coupling}
Among $\mathcal{L}_{\mu,\nu}$, some couplings $\pi$ convey more information onto the margins than others. We suppose we want to reduce the available information one can extract out of realizations from $\pi$, while picking $\pi$ as close to the uniform measure as possible. Applications for such an hypothesis can be found in \cite{bertrand2021minimization}.

A constant, necessarily the uniform law $\mathbb{U}^{pq}$ would convey not information. However, it obviously does not respect margins. Therefore, let us force $\pi$ to belong to $\mathcal{L}_{\mu,\nu}$ while being as close as possible to $\mathbb{U}^{pq}$. The square distance is a natural choice, actually motivated by the mean square error decomposition. We end up looking at:

\begin{problem}[Minimal Trade Model] \label{pb:chi2}
$\min_{\pi\in\mathcal{L}_{\mu,\nu}}\sum_{u=1}^p\sum_{v=1}^q\left(\pi_{u,v}-\mathbb{U}^{pq}_{u,v}\right)^2$
\end{problem}

It happens that we can compute the exact form of the solution (see~\cite{bertrandadac}). It is given by the so-called indeterminacy coupling $\pi^+$:

\begin{equation}\label{eq:def+}
\pi^+_{u,v} = (\mu\oplus\nu)_{u,v} = \frac{\mu_u}{q} + \frac{\nu_v}{p} - \frac{1}{pq},~\forall 1\le u \le p,~\forall 1\le v \le q
\end{equation}

This formula is positive if and only if the following inequality holds:
\begin{equation}
\label{cond:H}
\frac{\mu_0}{q} + \frac{\nu_0}{p} - \frac{1}{pq} \geq 0
\end{equation}
where $\mu_0 = \min_{\forall 1\le u \le p}\mu_u$ and $\nu_0 = \min_{\forall 1\le v \le q}\nu_u$.

Inequality~(\ref{cond:H}) considerably reduce the choice of the margins. In subsection~\ref{ssec:margins} we describe a projection which aims at changing any couple of margins into a couple of margins respecting Inequality~(\ref{cond:H}). Furthermore, we measure the proportion of margins eligible for an indeterminacy coupling. 

\subsection{Couple matchings minimization\label{ssec:collisions}}
The cost function in Problem~\ref{pb:chi2} leads to minimize:
\begin{equation}
\label{eq:dev_square}
\sum_{u=1}^p\sum_{v=1}^q\pi_{u,v}^2
\end{equation}

A first remark is that substituting $\mathbb{U}^{pq}_{u,v}$ by any constant in Problem~\ref{pb:chi2} would have led to the same simplification. Though, interpreting the constant as a probability measure requires its value to be $\frac{1}{pq}$.

\begin{mydef}[Couple Matching] \label{def:couple_matching}
Consider to independent draws $(U_1,V_1)$ and $(U_2,V_2)$ of $\pi$, a probability law in the simplex $S_{pq}$. A couple matching occurs when $U_1=U_2$ and $V_1=V_2$.
\end{mydef}

Equation~(\ref{eq:dev_square}) is nothing than probability that $\mathbb{P}((U_1,V_1) = (U_2,V_2))$, the probability of matching. Therefore $\pi^+$ minimizes the chances of couple matchings under any $\pi$ in $\mathcal{L}_{\mu,\nu}$.

\section{Properties of indeterminacy \label{sec:properties}}
\subsection{Measuring the subset of margins eligible for indeterminacy\label{ssec:margins}}

In this section, we estimate the impact of Inequality~(\ref{cond:H}) on the margins. We begin with a simple case: to construct $\pi^+$ coupling $\mu$ with itself, the pair ($\mu$,$\mu$) must satisfy~(\ref{cond:H}) which here writes $\mu_0 \ge \frac {1} {2p}.$ We estimate the probability that such an event happens. For this, consider the uniform distribution on $S_p$, the simplex of all laws on $p$ values and compute the normalized Lebesgue measure of the eligible subset of $S_p$.

\begin{prop} \label{pp:H_log2}
The proportion of $\mu$ in $S_p$ such that $(\mu,\mu)$ respects Inequality~(\ref{cond:H}) is $\frac{1}{2^{p-1}}$.
\end{prop}

\begin{proof}
By~(\ref{cond:H}) impose restricted bounds on the integrals constructing $\mu$. The eligible set has measure
\begin{eqnarray*}
\int_{\frac{1}{2p}}^{1-\frac{p-1}{2p}}
\int_{\frac{1}{2p}}^{1- \frac{p-2}{2p}-x_1}
\ldots
\int_{\frac{1}{2p}}^{1-\frac{1}{2p} - \sum_{i=1}^{p-2} x_i}
\diff x_1 \ldots \diff x_{p-1}. \\
\end{eqnarray*}

With the successive changes of variables $x_i \leftarrow x_i + \frac {1}{2p}$, this integral writes:
\begin{eqnarray*}
&&\int_{0}^{\frac{1}{2}}
\int_{0}^{\frac{1}{2} -x_1}
\ldots
\int_{0}^{\frac{1}{2} - \sum_{i = 1}^{p-2} x_i}.
\diff x_1 \ldots \diff x_{p-1} \\
&=& \frac{1}{2^{p-1}}
\int_{0}^{1}
\int_{0}^{1 -y_1}
\ldots
\int_{0}^{1 - \sum_{i = 1}^{p-2} y_i}.
\diff y_1 \ldots \diff y_{p-1} \\
&=& \frac{1}{2^{p-1}}.
\end{eqnarray*}
\end {proof}

\begin{rem} \label{req:constructionH}
The previous result is not surprising. A constructive method exists to build a valid $\mu$. Indeed, by Inequality~(\ref{cond:H}), $\mu_u$ is greater than $\frac{1}{2p}$ for all $u$. We deduce: $\mu_u=\frac{1}{2p} + \frac{r_u}{2}$ where $r$ is an arbitrary probability law on $p$ elements.
\end{rem} 

Following the same assumptions as before, we draw $\mu$ and $\nu$ uniformly and independently among the probability laws, therefore among $S_p$ and $S_q$ respectively. The proposition below characterizes eligible couples for an indeterminacy coupling.

\begin{prop} [Construction of eligible margins, discrete case]
\label{prop:alpha_discret}
~ \\
The couple of margins $(\mu, \nu)$ respects Inequality~(\ref{cond:H}) if and only if there exists a positive real $\alpha$ such that:

\begin{minipage}{0.4\textwidth}
    \begin{equation}
        \forall 1 \le u \le p,~\mu_u \geq \frac{\alpha}{p}
    \end{equation}
\end{minipage}
\hspace{4ex} % eventuellement
\begin{minipage}{0.4\textwidth}
    \begin{equation}
	    \forall 1 \le v \le q,~\nu_v \geq \frac{1-\alpha}{q}
    \end{equation}
\end{minipage}
\end{prop}
\begin{proof}
First, consider Inequality~(\ref{cond:H}) is satisfied and define defining $\alpha = p\mu_0 \in [0,1]$ then $\forall 1 \le v \le q$:
\begin{equation*}
	\frac{\mu_{0}}{q} + \frac{\nu_v}{p} \geq \frac{1}{pq}
\end{equation*}
which rewrites:
\begin{equation*}
	\nu_v \geq \frac{1-\alpha}{q}
\end{equation*}
Now, if such an $\alpha$ exists then $\forall 1 \le u \le p$ and $\forall 1 \le v \le q$, 
\begin{equation*}
\frac{\mu_u}{q} + \frac{\nu_v}{p} \geq \frac{\alpha}{pq} + \frac{1-\alpha}{pq} = \frac{1}{pq}
\end{equation*}
\end{proof}
\begin{rem}
Since $\mu_0$ is the minimum of a set of $p$ elements summing to $1$, $\alpha$ as well as $\beta = 1- \alpha$ belong to $[0,1]$. It symmetrically implies that $\nu_0\geq\frac{\beta}{q}$. 
\end{rem}
\begin{rem}
The introduction of the variable $p$ in the definition of $\alpha$ it saves the symmetry between $\mu$ and $\nu$ by ensuring that all the values $\alpha$ of $[0,1]$ are eligible regardless of $p$ or $q$. 
\end{rem}

Through a generalization of remark~\ref{req:constructionH}, Property~\ref{prop:alpha_discret} gives the existence of two probability laws $r$ and $s$ on $p$ and $q$ elements such as:

\begin{minipage}{0.4\textwidth}
    \begin{equation}
        \forall 1 \le u \le p,~\mu_u = \frac{\alpha}{p} + (1- \alpha)r_u 
        \label{eq:decompr}
    \end{equation}
\end{minipage}
\hspace{4ex} % eventuellement
\begin{minipage}{0.4\textwidth}
    \begin{equation}
	    \forall 1 \le v \le q,~\nu_v = \frac{1- \alpha}{q} + \alpha s_v 
	    \label{eq:decomps}
    \end{equation}
\end{minipage}

\begin{prop}[Constructive eligible margins]
A couple of probability laws $(\mu,\nu)\in S_p\times S_q$ respects Inequality~(\ref{cond:H}) if and only it it exists a real $\alpha\in[0,1]$ and a couple of probability laws $(r,s)\in S_p\times S_q$ such that Equations~(\ref{eq:decompr})~and~(\ref{eq:decomps}) are satisfied.
\end{prop}

For fixed $\alpha$, the the space of eligible $\mu$ appears as a $(1- \alpha)$-contraction of $S_p$ whereas that of $\nu$ is an $\alpha$-contraction of $S_q$. Since the two laws are drawn independently, the measure of the eligible space in $S_p \times S_q$ is given by:

\begin{equation}\label{eq:propmunu}
\int_{\alpha = 0} ^ 1 \alpha^{p-1}(1- \alpha)^{q-1} \diff\alpha = \frac{(p-1)!(q-1)!}{(p + q-2)!}
\end{equation}

The eligibility results are summarized in the following proposition:
\begin{prop} [Valid proportion] \label{prop: validProportion}
If $\mu$ is drawn in the simplex $S_p$ uniformly, the probability that the pair $(\mu, \mu)$ respects the Inequality~(\ref{cond:H}) is $\frac {1}{2^{p-1}}$. Then, there exists a probability law $r$ in the simplex $S_p$ such that $\mu$ satisfies:
\begin{equation}\label{eq: formMu}
\forall u, \mu_u = \frac{1}{2p} + \frac{r_u}{2}.
\end{equation}

If additionally $\nu$ is drawn in $S_q$, independently upon $\mu$ then, the probability that the pair $(\mu, \nu)$ respects Inequality(\ref{cond:H}) is $\frac{(p-1)!(q-1)!}{(p + q-2)!}$. In this case, there exists a real $\alpha$, a probability law $r$ in the simplex $S_p$ and a probability law  $s$ in the simplex $S_q$ such that:

\begin{minipage}{0.4\textwidth}
    \begin{equation}
		\forall u,~\mu_u = \frac{\alpha}{p} + (1- \alpha) r_u
		\label{eq:formMuCouple}
    \end{equation}
\end{minipage}
\hspace{4ex} % eventuellement
\begin{minipage}{0.4\textwidth}
    \begin{equation}
	    \forall v,~\nu_v = \frac{1- \alpha}{q} + \alpha s_v
	    \label{eq:formnuCouple}
    \end{equation}
\end{minipage}

In addition, the previous writings characterize compliance to Inequality~(\ref{cond:H}).
\end{prop}

\begin{rem}[Different shapes]
We notice that the expression of the eligible proportion depends on whether we are interested in the coupling of $\mu$ with itself or with a second and independent law $\nu$: the formula~(\ref{eq:propmunu}) of the second case does not catch up with the one in Property~\ref{pp:H_log2} by simply setting $p=q$. 
The difference comes from independency only holding in the second case.
\end{rem}

\subsection{Indeterminacy as a mixture of three independent couplings\label{ssec:decomposition}}

The formula which defines indeterminacy given in Equation~(\ref{eq:def+}) does not provide as such an efficient way to draw under indeterminacy nor any interpretation of the meaning of it. We propose to rewrite this formula so as to view indeterminacy as a classic mixture of three independent couplings. Our starting point is the usual form of an indeterminacy coupling.
\begin{equation*}
	\pi^+_{u,v} =\frac{\mu_u}{q} + \frac{\nu_v}{p} -\frac{1}{pq},~\forall 1\le u \le p,~\forall 1\le v \le q
\end{equation*}

Quoting $\mu_0 = \min_u \mu_u$ and $\nu_0 = \min_u \nu_u$ it rewrites:
\begin{equation*}
	\pi^+_{u,v} = \left[\frac{\mu_u-\mu_0}{q} \right] + \left[\frac{\nu_v-\nu_0}{p} \right] + \left[\frac{\mu_0}{q} + \frac{\nu_0}{p} - \frac{1}{pq}\right]
\end{equation*}

First let us remark that the three square brackets are positive since (\ref{cond:H}) is satisfied. Thus, we renormalize them to extract probability laws. Formally:
\begin{equation}\label{eq:decomp1}
	\pi^+_{u,v} = (1-p\mu_0)\left[\frac{\mu_u-\mu_0}{q(1-p\mu_0)} \right] + (1-q\nu_0)\left[\frac{\nu_v-\nu_0}{p(1-q\nu_0)} \right] + (p\mu_0 + q\nu_0 - 1)\left[\frac{1}{pq}\right]
\end{equation}

\begin{rem}[Tight case]
In case any of the two first brackets equals $0$ it means $\mu$ or $\nu$ is uniform. In that case indeterminacy and independence couplings are the same so that an interpretation of indeterminacy is trivial. 
Anticipating on the action of $T$ defined below, when equality holds in (\ref{cond:H}) then no uniform component exists leading to $R=3$ never happening.
\end{rem}

We now define a transformation $T$ acting on a probability law by:
\begin{mydef}\label{def:transfo}
Given a probability law $s=s_1,\ldots,s_r$ on $r$ elements, we quote $s_0$ its minimum. The transformation $T^r$ generates a new law on the same elements by:
\begin{eqnarray*}
	T^r&:& S_r\rightarrow S_r\\
	&&(s_i)_{1\le i \le r} \mapsto \left(\frac{s_i-s_0}{1-rs_0}\right)_{1\le i \le r}
\end{eqnarray*}
We shall quote $T$ the transformation acting on any $S_r$ through $T|_{S_r} = T^r$.
\end{mydef}
We notice that $T$ actually removes as much uniform part as possible from the probability law it operates on. $T(s)$ will concentrates its realizations on the modes of $s$. With this notation, Equation~(\ref{eq:decomp1}) rewrites:
\begin{equation}\label{eq:decompnorm}
	\pi^+_{u,v} = (1-p\mu_0)\frac{1}{q}T(\mu)_u + (1-q\nu_0)\frac{1}{p}T(\nu)_v + (p\mu_0 + q\nu_0 - 1)\mathbb{U}^{pq}_{u,v}
\end{equation}

Reading Equation~(\ref{eq:decompnorm}), we are able to decompose an indeterminacy draw as stated in proposition~\ref{prop:drawing+}.

\begin{prop}[Indeterminacy drawing decomposition]\label{prop:drawing+}
We introduce a random variable $R$ on $3$ modalities $1,2,3$ with respective probabilities $1-p\mu_0$, $1-q\nu_0$ and $p\mu_0 + q\nu_0 -1$. Realizations under indeterminacy eventually decomposes as a mixture of three straightforward drawings:
\begin{enumerate}
\item draw $R$;
\item if $R=1$ then $(u,v)$ is drawn under the independence coupling of $T(\mu)$ and $\mathbb{U}^q$;
\item if $R=2$ then $(u,v)$ is drawn under the independence coupling of $\mathbb{U}^p$ and $T(\mu)$;
\item if $R=3$ then $(u,v)$ is drawn under the independence coupling of $\mathbb{U}^p$ and $\mathbb{U}^q$ (\textit{i.e.} $\mathbb{U}^{pq}$).
\end{enumerate}
\end{prop}

Under this form, it appears that $\pi^+$ exhausts the uniform part of each margin. It is definitely consistent with indeterminacy being the projection of $\mathbb{U}^{pq}$ on $\mathcal{L}_{\mu,\nu}$. $T(\mu)$ is more concentrated on the modes of $\mu$ than $\mu$ itself. Consequently when $R=1$, $U$ is concentrated on the mode of $\mu$, far from the uniform: this is the price for the margin being $\mu$. On any other value of $R$, $U$ is uniformly drawn. Symmetrically, for $V$, the concentration on modes of $\nu$ happens when $R=2$. 

Eventually, Proposition~\ref{prop:drawing+} justifies the method induced by indeterminacy to reduce couple matchings. If $R=1$ a couple matching is rare since $U_1=U_2$ is prevented by $U$ being drawn uniformly under $U^p$; if $R=2$ then $V$ is drawn uniformly; if $R=3$ then both are drawn uniformly.

\section{Janson Vegelius coefficient\label{sec:jv}}

In statistical analysis, given independent realizations $(U_1,V_1),\ldots,(U_n,V_n)$, the categorization of $n$ individuals under two measures (for instance the city they live in, their socio-professional category, their ages, \ldots), how do we measure the correlation between $U$ and $V$? A solution is to use a deviation-to-independence coefficient, for instance: the $\chi_2$. To compute its value, from the $n$ realizations of $(U,V)$, we deduce an empirical margin $\pi$ counting the proportion of individuals in each couple of modalities:

\begin{equation}\label{eq:integerform}
\pi_{u,v} = \frac{\#\{i~/~U_i=u~\&~V_i=v\}}{n},~\forall 1\le u \le p,~\forall 1\le v \le q;
\end{equation}
similarly, denote $\mu$ and $\nu$ the empirical margins. The empirical $\chi^2$ index, denoted $\chi^2_n$ is then defined:
\begin{equation}
\label{eq:chi2}
\chi^2_n(U,V) = \sum_{u=1}^p\sum_{v=1}^q \frac{\left(\pi_{u,v}-(\mu\otimes\nu)_{u,v}\right)^2}{(\mu\otimes\nu)_{u,v}};
\end{equation}
which obviously happens to be null if and only if the empirical distribution $\pi$ of the observed data is an independence coupling of the empiric margins. Obviously, such an event never happens even under independence.

Using a symmetric idea, a lesser known criterion, called Janson-Vegelius Index, after the name of the inventors of this coefficient (see \cite{JV77}, \cite{JV78} or \cite{JV82}) writes as a deviation to indeterminacy:
\begin{equation}
\label{eq:jv}
JV_n(U,V) = \sum_{u=1}^p\sum_{v=1}^q \frac{\left(\pi_{u,v}-(\mu\oplus\nu)_{u,v}\right)^2}{\sqrt{\frac{p-2}{p}\left(\sum_{u=1}^q\mu_u^2 + 1\right)}\sqrt{\frac{q-2}{q}\left(\sum_{v=1}^q\nu_v^2 + 1\right)}};
\end{equation}
and obviously is equal to zero if and only if the empirical $\pi$ is an indeterminacy coupling of the empirical margins as defined in Equation~(\ref{eq:def+}). We omit the subscript $n$ in the following. The $JV$ index is actually just a classical cosine coefficient when rewritten in the "Mathematical Relational Analysis" Space. The relational analysis space no longer encodes modalities but links between individuals. Two matrices $X$ and $Y$ of size $n\times n$ respectively associated to variables $U$ and $V$ are introduced as shown in Definition~\ref{def:arm_notations}.

\begin{mydef}[Mathematical Relational Analysis notations]
\label{def:arm_notations}
Let $(U_1,\ldots,U_n)$ and $(V_1,\ldots,V_n)$ be two $n$ probabilistic draws of $U$ and $V$ respectively.
We define two associated symmetric $n \times n$ matrices $X$ and $Y$ by:

\begin{minipage}{0.4\textwidth}
    \begin{equation}
        X_{i,j} = \mathds{1}_{U_i=U_j},~ \forall 1\le i,j \le n
    \end{equation}
\end{minipage}
\hspace{4ex} % eventuellement
\begin{minipage}{0.4\textwidth}
    \begin{equation}
	    Y_{i,j} = \mathds{1}_{V_i=V_j},~ \forall 1\le i,j \le n
    \end{equation}
\end{minipage}
\end{mydef}

To understand the notation, let us begin with some remarks about Definition~\ref{def:arm_notations}. Basically, the two $\{0,1\}$ matrices $X$ and $Y$ represent agreements and disagreements between the two variables on a same draw of size $n$; they are symmetric with $1$ values on their diagonal. As expected, one can pass from the relational encoding to the usual contingency encoding as well as in the reciprocal way; those transfer formulas are demonstrated in the mentioned articles and enable us to write $JV$ as a cosine in the relational space:

\begin{equation}
\label{eq:jvrel}
JV(U,V) = JV(X,Y) = \frac{\sum_{i=1}^n\sum_{j=1}^n\left(X_{i,j}-\frac{1}{p}\right)\left(Y_{i,j}-\frac{1}{q}\right)}{\sqrt{\sum_{i=1}^n\sum_{j=1}^n\left(X_{i,j}-\frac{1}{p}\right)^2\sum_{i=1}^n\sum_{j=1}^n\left(Y_{i,j}-\frac{1}{q}\right)^2}}
\end{equation}

Calculations leading to Equation~(\ref{eq:jvrel}) from Equation~(\ref{eq:jv}) can be found in \cite{MAM79} or \cite{MAY91}. Additionally, key features about relational analysis can be found in \cite{MAR84}, \cite{MES90}, \cite{Opitz05}, \cite{MAR86}, \cite{MAR91} and \cite{AHP09a}.

\subsection{Average value of JV through simulation}
\begin{figure}[H]
\begin{center}
    \includegraphics [trim = 0cm 0cm 0cm 1.2cm, clip, scale = 0.35]{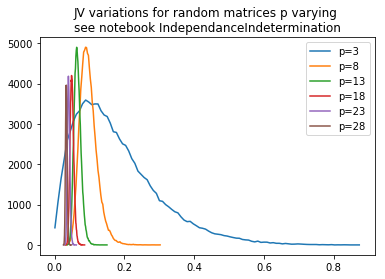}
	\caption{Distribution of the $JV$ when $\pi$ is uniform in $S_{p^2}$ 
	\label{fig: JVSim}}
\end{center}
\end{figure}

We simulate random probability laws $\pi$ uniformly in $S_{p^2}$ to compute the distribution of the criterion $JV$  according to $p$. We first propose Figure~\ref{fig: JVSim} which presents the distribution of the criterion. One element strikes immediately: values concentrate around $0$ as $p$ increases. We start proving it in the case $\pi=\mu\otimes\mu$ for which the formula is simplified before demonstrating the general case.

\subsection{Average value of JV through computations}
This section gives a proof (in the case of the independent coupling of $\mu$ with itself) of the limit property noted in the previous section. It is stated in Proposition~\ref{th:JV0}.
\begin{prop}[JV limit, independence case]
\label{th:JV0}
~\\
If $\mu$ is uniform in $S_p$ then $\lim_{p \rightarrow \infty} \mathbb{E} _{\mu} \left [JV (\mu \otimes \mu) \right] = 0 $
\end{prop}

\begin{proof}
We start by using a sequence of inequalities to extract an upper bound of $JV$.
\begin{eqnarray}
JV (\mu \otimes \mu) & = & \frac{p ^ 2 \sum_{u = 1} ^ p{\left (\mu_u- \frac{1}{p} \right) ^ 2} \sum_{u = 1} ^ p{\left (\mu_u- \frac{1}{p} \right) ^ 2}}{p (p-2) (\sum_{u = 1} ^ p{\mu_u ^ 2}) + 1} \nonumber\\
& = & \frac{p ^ 2 \left (\sum_{u = 1} ^ p{\mu_u ^ 2- \frac{1}{p}} \right) ^ 2}{p (p-2) ( \sum_{u = 1} ^ p{\mu_u ^ 2}) + 1} 
\le \frac{p ^ 2 \left (\sum_{u = 1} ^ p{\mu_u ^ 2- \frac{1}{p}} \right) ^ 2}{p ^ 2 (p- 2) \frac{1}{p ^ 2} +1} 
= \frac{p ^ 2 \left (\sum_{u = 1} ^ p{\mu_u ^ 2- \frac{1}{p}} \right) ^ 2}{p-1} \nonumber\\
&\le& 2p \left (\sum_{u = 1} ^ p{\mu_u ^ 2- \frac{1}{p}} \right) ^ 2 \label{eq:majdirichlet1}
\end{eqnarray}

To demonstrate the convergence, we introduce Dirichlet law:

\begin{mydef}[Dirichlet law]
\label{def:DirichletLaw}
The density of the Dirichlet law of parameter $(\alpha_1,\ldots,\alpha_p)\in(\mathbb{R}^{+*})^p$ on the simplex $S_p$ is expressed as follows:
$$ f(\mu_1, \ldots, \mu_p, \alpha_1,\ldots,\alpha_p)\prod_{k = 1}^ p{\diff\mu_k} = \frac{1}{B(\alpha)}\prod_{u = 1}^p\mu_u^{\alpha_u-1}\prod_{u = 1}^p {\diff\mu_u} \mathds{1}_{(\mu_1,\ldots,\mu_p) \in S_p} $$
where, if we denote $\Gamma$ the usual gamma function, $B$ is the mutinomial beta function:
\begin{equation*}
B(\alpha) = \frac{\prod_{u=1}^p\Gamma(\alpha_u)}{\Gamma\left(\sum_{u=1}^p\alpha_u\right)}.
\end{equation*}

The particular case when $\alpha_u=1,~\forall 1\le u \le p$ expresses a uniform law on $S_p$ whose density is given by:
$$ f(\mu_1, ..., \mu_p)\prod_{k = 1}^ p{\diff\mu_k} = (p-1)! \prod_{k = 1}^p {\diff\mu_k} \mathds{1}_{(\mu_1,\ldots,\mu_p) \in S_p} $$

\end{mydef}

We specify the moments of the Dirichlet law to deduce the exact calculation of the expectation of the upper bound. 

\begin{prop} [Dirichlet law moments]
\label{prop: mom_dirichlet}
Given $\mu\in S_p$ drawn according to Dirichlet law with parameter $(\alpha_1, \ldots, \alpha_p)$. Denote $ \alpha_0 = \sum_{u=1}^p \alpha_{u} $. Then for all $p$-uplet $ \beta_1, \ldots, \beta_p $ of positive integers, we have the formula (with $\beta_0 = \sum_{u=1}^p \beta_{u}$):
$$ \mathbb{E} \left (\prod_{u = 1} ^ p \mu_u ^{\beta_u} \right) = \frac{\Gamma \left (\sum_{u = 1} ^ p \alpha_u \right)}{\Gamma \left (\sum_{u = 1} ^ p \alpha_u + \beta_u \right)} \prod_{u = 1} ^ p \frac{\Gamma (\alpha_u + \beta_u)}{\Gamma (\alpha_u)} = \frac{\Gamma \left (\alpha_0 \right)}{\Gamma \left (\alpha_0 + \beta_0 \right)} \prod_{u = 1} ^ p \frac{\Gamma (\alpha_u + \beta_u)}{\Gamma (\alpha_u)} $$
\end{prop}

We now develop the upper-bound and we evaluate all terms making use of Proposition~\ref{prop: mom_dirichlet} with $ \alpha_1 = \ldots = \alpha_p = 1 $.
\begin{equation*}
\mathbb{E} \left [\left (\sum_{u = 1} ^ p{\mu_u ^ 2- \frac{1}{p}} \right) ^ 2 \right] = \sum_{1 \le u, v \le p} \mathbb{E} (\mu_u ^ 2 \mu_v ^ 2) - \frac{2}{p} \sum_{u = 1} ^ p \mathbb{E} (\mu_u ^ 2 ) + \frac{1}{p ^ 2}
\end{equation*}

For any $ u \neq v $, it holds
\begin{equation}
\mathbb{E} (\mu_u ^ 2 \mu_v ^ 2) = \frac{\Gamma (\alpha_0)}{\Gamma (\alpha_0 + \beta_0)} \frac{\Gamma (\alpha_u + 2) \Gamma (\alpha_v + 2)}{\Gamma (\alpha_u) \Gamma (\alpha_v)} = \frac{4}{p (p + 1) (p + 2) (p + 3)} \label{eq: esp_22}.
\end{equation}
For $u=v$ then
\begin{equation}
\mathbb{E} (\mu_u ^ 4) = \frac{2 * 3 * 4}{p (p + 1) (p + 2) (p + 3)} = \frac{24}{p (p + 1) (p + 2) (p + 3)} \label{eq: esp_4}.
\end{equation}
Finally,
\begin{eqnarray}
\mathbb{E} (\mu_u ^ 2) & = & \frac{2}{p (p + 1)} \label{eq: esp_2}
\end{eqnarray}
Summing up, by Equations~(\ref{eq: esp_22}), (\ref{eq: esp_4}) and (\ref{eq: esp_2}), we get:
\begin{eqnarray*}
\mathbb{E} \left [\left (\sum_{u = 1} ^ p{\mu_u ^ 2- \frac{1}{p}} \right) ^ 2 \right] & = & \frac{4p (p-1)}{p (p + 1) (p + 2) (p + 3)} + \frac{24p}{p (p + 1) (p + 2) (p + 3)} - \frac{2 * 2p}{p (p + 1)} + \frac{1}{p ^ 2} \\
& = & \frac{4 (p-1)}{(p + 1) (p + 2) (p + 3)} + \frac{24}{(p + 1) (p + 2) (p + 3)} - \frac{4p}{(p + 1)} + \frac{1}{p ^ 2} \\
& = & \frac{4p (p + 5) -4 (p + 2) (p + 3)}{p (p + 1) (p + 2) (p + 3)} + \frac{1}{p ^ 2} = \frac{-24}{p (p + 1) (p + 2) (p + 3)} + \frac{1}{p ^ 2}
\end{eqnarray*}
Therefore by Inequality~(\ref{eq:majdirichlet1}) we know that $\mathbb{E}_{\mu} \left [JV (\mu \otimes \mu) \right] = \mathcal{O} (\frac{1}{p}) $ which ends the proof of Proposition~\ref{th:JV0}.
\end{proof}

We now prove the following more general result.

\begin{prop} [JV limit] ~ \\
\label{th:JV0_gen}
If $\pi$ is uniform in $S_{pq}$ then $ \lim_{pq \rightarrow \infty} \mathbb{E} _{\pi} \left [JV (\pi) \right] = 0 $
\end{prop}
\begin{proof}
For all $1\le u\le p$ and all $1\le v \le q$, we denote $\pi_{u,\cdot} = \sum_{v=1}^q\pi_{u,v}$ and $\pi_{\cdot,v} = \sum_{u=1}^p\pi_{u,v}$.
As in the case when $\pi=\mu\otimes\mu$, it holds:
\begin{equation*}
JV (\pi) \le \frac{4}{\sqrt{pq}} \left (pq \sum_{u, v} (\pi_{u, v} ^ 2) -p \sum_{u} (\pi_{u, \cdot} ^ 2) -q \sum_{v} (\pi _{\cdot, v} ^ 2) +1 \right)
\end{equation*}
We evaluate the expectation of the above upper bound making use of Proposition~\ref{prop: mom_dirichlet} with a uniform Dirichlet law on the simplex $S_{pq}$. Therefore consider $ (\pi_{u, v})_{1 \le u \le p, 1 \le v \le q} $ a vector of size $ pq $. Then, for the first term, $\mathbb{E} (\pi_{u, v} ^ 2) = \frac{2}{pq (pq + 1)}$ and for the second term,
\begin{equation}
\mathbb{E} (\pi_{u, \cdot} ^ 2) = \mathbb{E} \left (\sum_{v, v '} \pi_{u, v} \pi_{u, v'} \right) = (q ^ 2-q) \frac{1}{pq (pq + 1)} + q \frac{2}{pq (pq + 1)} = \frac{q + 1}{p (pq + 1)} \label{eq: esp_2_point}
\end{equation}

By combining both equations and playing with symmetry for the third term, we obtain:
\begin{eqnarray*}
& \mathbb{E} & \left (pq \sum_{u, v} (\pi_{u, v} ^ 2) -p \sum_{u} (\pi_{u, \cdot} ^ 2) -q \sum_{v} (\pi _{\cdot, v} ^ 2) +1 \right) \\
&& = p ^ 2q ^ 2 \frac{2}{pq (pq + 1)} -p ^ 2 \frac{q + 1}{p (pq + 1)} -q ^ 2 \frac{p + 1}{q (pq + 1)} +1 \\
&& = \frac{2pq}{pq + 1} - \frac{p (q + 1)}{pq + 1} - \frac{q (p + 1)}{pq + 1} +1 = \frac{-p-q}{pq + 1} +1.
\end{eqnarray*}
Eventually, $\mathbb{E} _{\pi} \left [JV (\pi) \right] = \mathcal{O} \left (\frac{1}{\sqrt{pq}} \right)$.
\end{proof}

\begin{rem}
Since $JV$ is positive we have also shown that it tends towards $0$ in probability. Besides, The propensity of the JV to approach $0$ assumes that we use it sparingly before assuming indeterminacy.
\end{rem}

\section{Conclusion}
The main innovation of this paper is the decomposition of indeterminacy. It enables us, first to efficiently generate a drawing, second to interpret it as a mixture of three straightforward drawings and last but not least to explain how it reduces couple matchings while respecting the forced margins. 
Since indeterminacy cannot be defined on all margins, the paper also computes the proportion of eligible margins. Furthermore, it proposes a constructive method to transform any couple into an eligible couple. Eventually, the 0-limit of the Janson Vegelius coefficient helps us to mind when defining a threshold to conclude to indeterminacy. 

\bibliographystyle{elsarticle-harv}
\bibliography{BiblioSPL}

\begin{thebibliography}{16}
\expandafter\ifx\csname natexlab\endcsname\relax\def\natexlab#1{#1}\fi
\providecommand{\url}[1]{\texttt{#1}}
\providecommand{\href}[2]{#2}
\providecommand{\path}[1]{#1}
\providecommand{\DOIprefix}{doi:}
\providecommand{\ArXivprefix}{arXiv:}
\providecommand{\URLprefix}{URL: }
\providecommand{\Pubmedprefix}{pmid:}
\providecommand{\doi}[1]{\href{http://dx.doi.org/#1}{\path{#1}}}
\providecommand{\Pubmed}[1]{\href{pmid:#1}{\path{#1}}}
\providecommand{\bibinfo}[2]{#2}
\ifx\xfnm\relax \def\xfnm[#1]{\unskip,\space#1}\fi
%Type = Inproceedings
\bibitem[{Ah-Pine(2010)}]{AHP09a}
\bibinfo{author}{Ah-Pine, J.}, \bibinfo{year}{2010}.
\newblock \bibinfo{title}{On aggregating binary relations using 0-1 integer
  linear programming}, in: \bibinfo{booktitle}{ISAIM}, pp.
  \bibinfo{pages}{1--10}.
%Type = Inproceedings
\bibitem[{Bertrand et~al.(2021)Bertrand, Broniatowski and
  Marcotorchino}]{bertrand2021minimization}
\bibinfo{author}{Bertrand, P.}, \bibinfo{author}{Broniatowski, M.},
  \bibinfo{author}{Marcotorchino, J.F.}, \bibinfo{year}{2021}.
\newblock \bibinfo{title}{Minimization with respect to divergences and
  applications}, in: \bibinfo{booktitle}{International Conference on Geometric
  Science of Information}, \bibinfo{organization}{Springer}. pp.
  \bibinfo{pages}{818--828}.
%Type = Article
\bibitem[{Bertrand et~al.(2022)Bertrand, Broniatowski and
  Marcotorchino}]{bertrandadac}
\bibinfo{author}{Bertrand, P.}, \bibinfo{author}{Broniatowski, M.},
  \bibinfo{author}{Marcotorchino, J.F.}, \bibinfo{year}{2022}.
\newblock \bibinfo{title}{Independence versus indetermination: basis of two
  canonical clustering criteria}.
\newblock \bibinfo{journal}{Advances in Data Analysis and Classification} ,
  \bibinfo{pages}{1--25}.
%Type = Phdthesis
\bibitem[{Conde-C{\'e}spedes(2013)}]{PCThese}
\bibinfo{author}{Conde-C{\'e}spedes, P.}, \bibinfo{year}{2013}.
\newblock \bibinfo{title}{Modélisations et extensions du formalisme de
  l'analyse relationnelle mathématique à la modularisation des grands
  graphes}.
\newblock Ph.D. thesis. Paris 6.
%Type = Article
\bibitem[{Csisz{\'a}r et~al.(1991)}]{csiszar1991least}
\bibinfo{author}{Csisz{\'a}r, I.}, et~al., \bibinfo{year}{1991}.
\newblock \bibinfo{title}{Why least squares and maximum entropy? an axiomatic
  approach to inference for linear inverse problems}.
\newblock \bibinfo{journal}{The annals of statistics} \bibinfo{volume}{19},
  \bibinfo{pages}{2032--2066}.
%Type = Article
\bibitem[{Janson and Vegelius(1977)}]{JV77}
\bibinfo{author}{Janson, S.}, \bibinfo{author}{Vegelius, J.},
  \bibinfo{year}{1977}.
\newblock \bibinfo{title}{Correlation coefficients for nominal scales}.
\newblock \bibinfo{journal}{Uppsala: Department of Statistics} .
%Type = Article
\bibitem[{Janson and Vegelius(1978)}]{JV78}
\bibinfo{author}{Janson, S.}, \bibinfo{author}{Vegelius, J.},
  \bibinfo{year}{1978}.
\newblock \bibinfo{title}{On the applicability of truncated component analysis
  based on correlation coefficients for nominal scales}.
\newblock \bibinfo{journal}{Applied Psychological Measurement} ,
  \bibinfo{pages}{135--145}.
%Type = Article
\bibitem[{Janson and Vegelius(1982)}]{JV82}
\bibinfo{author}{Janson, S.}, \bibinfo{author}{Vegelius, J.},
  \bibinfo{year}{1982}.
\newblock \bibinfo{title}{The {J}-index as a measure of nominal scale response
  agreement}.
\newblock \bibinfo{journal}{Applied Psychological Measurement} ,
  \bibinfo{pages}{111--121}.
%Type = Article
\bibitem[{Marcotorchino(1984)}]{MAR84}
\bibinfo{author}{Marcotorchino, J.F.}, \bibinfo{year}{1984}.
\newblock \bibinfo{title}{Utilisation des comparaisons par paires en
  statistique des contingences}.
\newblock \bibinfo{journal}{Publication du Centre Scientifique IBM de Paris et
  Cahiers du Séminaire Analyse des Données et Processus Stochastiques
  Université Libre de Bruxelles} , \bibinfo{pages}{1--57}.
%Type = Article
\bibitem[{Marcotorchino(1986)}]{MAR86}
\bibinfo{author}{Marcotorchino, J.F.}, \bibinfo{year}{1986}.
\newblock \bibinfo{title}{Maximal association theory as a tool of research}.
\newblock \bibinfo{journal}{Classification as a tool of research , W.Gaul and
  M. Schader editors, North Holland Amsterdam} .
%Type = Article
\bibitem[{Marcotorchino(1991)}]{MAR91}
\bibinfo{author}{Marcotorchino, J.F.}, \bibinfo{year}{1991}.
\newblock \bibinfo{title}{Seriation problems:an overview}.
\newblock \bibinfo{journal}{Applied Stochastic Models and Data Analysis}
  \bibinfo{volume}{7}, \bibinfo{pages}{139--151}.
%Type = Article
\bibitem[{Marcotorchino and El~Ayoubi(1991)}]{MAY91}
\bibinfo{author}{Marcotorchino, J.F.}, \bibinfo{author}{El~Ayoubi, N.},
  \bibinfo{year}{1991}.
\newblock \bibinfo{title}{Paradigme logique des écritures relationnelles de
  quelques critères fondamentaux d'association}.
\newblock \bibinfo{journal}{Revue de Statistique Appliquée}
  \bibinfo{volume}{39}, \bibinfo{pages}{25--46}.
%Type = Book
\bibitem[{Marcotorchino and Michaud(1979)}]{MAM79}
\bibinfo{author}{Marcotorchino, J.F.}, \bibinfo{author}{Michaud, P.},
  \bibinfo{year}{1979}.
\newblock \bibinfo{title}{Optimisation en Analyse Ordinale des Données}.
\newblock \bibinfo{publisher}{Ed Masson, Paris}.
%Type = Article
\bibitem[{Messatfa(1990)}]{MES90}
\bibinfo{author}{Messatfa, H.}, \bibinfo{year}{1990}.
\newblock \bibinfo{title}{Maximal association for the sum of squares of a
  contingency table}.
\newblock \bibinfo{journal}{Revue RAIRO, Recherche Opérationnelle}
  \bibinfo{volume}{24}, \bibinfo{pages}{29--47}.
%Type = Article
\bibitem[{Opitz and Paul(2005)}]{Opitz05}
\bibinfo{author}{Opitz, O.}, \bibinfo{author}{Paul, H.}, \bibinfo{year}{2005}.
\newblock \bibinfo{title}{Aggregation of ordinal judgements based on
  condorcet{’}s majority rule}.
\newblock \bibinfo{journal}{Data Analysis and Decision Support. Studies in
  Classification, Data Analysis, and Knowledge Organization. Springer, Berlin,
  Heidelberg} .
%Type = Article
\bibitem[{Sklar(1973)}]{Sklar73}
\bibinfo{author}{Sklar, A.}, \bibinfo{year}{1973}.
\newblock \bibinfo{title}{Random variables, joint distribution functions, and
  copulas}.
\newblock \bibinfo{journal}{Kybernetika} \bibinfo{volume}{9},
  \bibinfo{pages}{449--460}.

\end{thebibliography}

\end{document}